\documentclass[1p]{elsarticle}

\def\be{\begin{equation}}
\def\ee{\end{equation}}
\def\ba{\begin{eqnarray}}
\def\ea{\end{eqnarray}}
\def\bas{\begin{eqnarray*}}
\def\eas{\end{eqnarray*}}

\begin{document}

\begin{frontmatter}

\title{Effective theory for deformed nuclei}

\author{T.~Papenbrock}

\ead{tpapenbr@utk.edu}

\address{Department of Physics and Astronomy, University of
  Tennessee, Knoxville, TN 37996, USA}

\address{Physics Division, Oak Ridge National Laboratory, Oak
  Ridge, TN 37831, USA}

\address {GSI Helmholtzzentrum f\"ur Schwerionenforschung GmbH,
  64291 Darmstadt, Germany}

\address{Institut f\"ur Kernphysik, Technische Universit\"at
  Darmstadt, 64289 Darmstadt, Germany}

\begin{abstract}
  Techniques from effective field theory are applied to nuclear
  rotation. This approach exploits the spontaneous breaking of
  rotational symmetry and the separation of scale between low-energy
  Nambu-Goldstone rotational modes and high-energy vibrational and
  nucleonic degrees of freedom. A power counting is established and
  the Hamiltonian is constructed at next-to-leading order.
\end{abstract}

%\pacs{21.60.Ev,21.30.Fe,21.10.Re,33.15.Mt}
%\maketitle 
\end{frontmatter}
\section{Introduction}

Collective states are the low-lying excitations of heavy nuclei.  
One of the hallmarks of deformed nuclei are rotational bands
of the form
\be 
\label{master}
E(l)\approx A[l(l+1)-I_0(I_0+1)] \ .  \ee Here, $l\ge I_0$ denotes the
angular momentum, $A$ is a constant determined by fit to data, and
$I_0$ is the spin of the band head under consideration.  Rotational
states are the lowest-lying excitations in open-shell heavy nuclei. In
rotational even-even nuclei in the rare-earth region, for instance,
the $l=2$ state of the ($I_0=0$) ground-state band has an excitation
energy of about 80~keV while the vibrational $I_0=2$ band head is at
about 1~MeV of excitation energy.  Our understanding of nuclear
rotation is based on the ground-breaking papers by Bohr~\cite{Bohr},
and by Bohr and Mottelson~\cite{BM}. Here, collective nuclear
excitations are modeled in terms of quadrupole vibrations of the
nuclear surface.  Rotational nuclei are intrinsically deformed, i.e.
they exhibit a mostly axially symmetric static deformation in the
co-rotating intrinsic reference frame, giving rise to rotational bands
of the form of Eq.~(\ref{master}). The Bohr-Mottelson model has been
extended to include nuclei with tri-axial deformation within the
asymmetric rotor model~\cite{Da58}. Rotational nuclear states are also
described within the variable moment of inertia
model~\cite{ScharffGoldhaber}, and the general collective
model~\cite{Gneuss,Hess1980,Greiner}.  The interacting boson
model~\cite{Arima}, an algebraic model based on $s$-wave and $d$-wave
bosons, is also widely employed for the description of collective
nuclear phenomena.  While these models are very appealing due to their
physical motivation and due to their mathematical beauty, it is
difficult to systematically extend them, to gauge their limitations,
or to compute results with reliable error estimates.  Furthermore, it
is non-trivial to keep generalizations of collective models
computationally tractable~\cite{Rowe,Caprio}. This is due to the
difficulties posed by the linear realization of rotational symmetry.

The present paper attempts to overcome these limitations. It proposes
a model-independent description of rotational nuclei that is based on
an effective theory (EFT). It treats even-even, odd mass and odd-odd
nuclei on an equal footing.  In recent years, EFTs enjoyed considerable
popularity and success in low-energy nuclear structure. Examples are
the application of pion-less EFT to few-body systems~\cite{KSW},
dilute Fermi gases with repulsive~\cite{Furnstahl} and attractive
\cite{Marini,PapBertsch,Hammer} interactions, nuclear interactions
based on chiral EFT~\cite{vKolck,Epelbaum,Machleidt}, and the
description of halo and cluster states within an
EFT~\cite{Bertulani,Riga}.  Effective field theories exploit a
separation of scale, and provide us with a model independent
description of physical phenomena based on
observables~\cite{DBKaplan,Burgess}. They often exhibit an impressive
efficiency as highlighted by analytical results and economical means
of calculations.

Within an effective theory, one proceeds as follows.  First, the
relevant low-energy degrees of freedom have to be identified. For
even-even nuclei in the rare earth region, for instance, the spins and
parities of low-lying states can be explained in terms of quadrupole
degrees of freedom.  Second, the relevant symmetries (and pattern of
spontaneous symmetry breaking) have to be identified.  For atomic
nuclei, the Hamiltonian is invariant under rotations.  Furthermore,
the concept of an intrinsically deformed ground state corresponds to
the spontaneous breaking of rotational symmetry. (See Bohr's Nobel
lecture~\cite{BohrNobel}, or Mottelson's lectures in Les
Houches~\cite{BenLesHouches}.)  Indeed, the rotational
spectrum~(\ref{master}) consists of the low-lying excitations
associated with the spontaneous breakdown of rotational symmetry. In
an infinite system, the corresponding excitations are massless
Nambu-Goldstone bosons. In a finite system such as the atomic nucleus,
there is -- strictly speaking -- no spontaneous symmetry breaking and
thus no Nambu-Goldstone boson. However, as we will see below, the
Nambu-Goldstone modes generate the low-energetic discrete
excitations~(\ref{master}) upon the quantization of the finite
system~\cite{Ui1983}.  Third, we have to identify (and exploit) a
separation of scales and introduce a power counting.  In the case of
deformed even-even nuclei in the rare earth region, the separation of
scales is the separation between, e.g. the rotational $J^\pi=2^+$
state (at several tens of keV of energy) and the low-lying vibrational
$2^+$ state (at about 1~MeV of excitation energy).  Below the
vibrational threshold, the physics can be described purely within
Nambu-Goldstone modes.  Above the threshold, the vibrations have
explicitly to be taken into account, and one might also consider the
coupling to even higher energetic nucleonic degrees of freedom (with a
corresponding energy scale of a few to several MeV).  For even-even
nuclei, the phenomenological
models~\cite{BM,Gneuss,Hess1980,Greiner,Arima} practically include the
physics of the Nambu-Goldstone modes and the vibrational degrees of
freedom.  However, the models do not further exploit the separation of
scales, and they do not present a power counting that would allow for
a systematic extension.

We want to apply the tools of effective field theory to
nuclear rotation. The concept of spontaneous symmetry breaking is
central to this approach. The ground breaking papers by
Weinberg~\cite{Weinberg}, by Coleman, Wess and Zumino~\cite{CWZ}, and
by Callan, Coleman, Wess and Zumino~\cite{CCWZ} describe the
construction of low-energy Lagrangians in the presence of spontaneous
symmetry breaking. Leutwyler applied and extended this approach to
non-relativistic Lagrangians~\cite{Leut93}, and excellent description
of the approach can be found in
Refs.~\cite{Weinberg2,Para,Miransky,Burgess,Brauner}. As we will see
below, the Nambu-Goldstone fields employed in the spontaneous
breakdown of rotational symmetry only depend on time (and not on
space).  This is due the finite size of the atomic nucleus. Thus, we
do not deal with a field theory but rather with quantum mechanics.
For this reason, this approach is an effective theory
(and not an EFT). Note that nuclear rotation and the
spontaneous breakdown of rotational symmetry has been addressed within
a field theoretical approach by Fujikawa and Ui~\cite{Ui1986}. These
authors succeeded in linking nuclear rotation to the Higgs mechanism,
but they did not pursue the systematic construction of low-energy
Lagrangians.

There are -- of course -- microscopic approaches to nuclear rotation
(see the recent review~\cite{Wyss} and references therein), and the
microscopic computation of the parameters of collective models is a
long-standing~\cite{Villars} and interesting
problem~\cite{Bender,Niksic,Nomura,Roh}. The aim of the present paper
is not to extract a collective model from an underlying microscopic
Hamiltonian, but rather to construct a low-energy collective
Hamiltonian within an effective theory. Apart from symmetry
principles, no details of the microscopic Hamiltonian are needed for
this task.

This paper is organized as follows. Section~\ref{sec:coset} presents
the derivation of the low-energy Lagrangians and Hamiltonians that
govern the dynamics of axially deformed nuclei, i.e. the spontaneous
breakdown of $SO(3)$ to $SO(2)$ is the central concept. In
Sect.~\ref{sec:NG} we consider the physics of Nambu-Goldstone modes
that result from the symmetry breaking. In Sect.~\ref{sec:vib} we
study the coupling of vibrations to the Nambu-Goldstone modes.
Fermions and Nambu-Goldstone modes are coupled in
Sect.~\ref{sec:fermions}. The main results of this paper are summarized in
Sect.~\ref{sec:summary}. Some technical details are presented in the
Appendix.

\section{Construction of low-energy Lagrangians}
\label{sec:coset}

Let us consider a system with a continuous symmetry such as invariance
under rotations.  In this case, the ground state has definite angular
momentum.  Spontaneous symmetry breaking happens when an {\it
  arbitrarily small} symmetry-breaking perturbation 
yields a deformed ground state, i.e. a state with no definite angular momentum
(see, e.g., Ref.~\cite{Wezel}). Clearly, spontaneous symmetry breaking
can only take place in infinite systems as only these can exhibit
gap-less excitations.

Ferromagnets and antiferromagnets are well known examples for the
spontaneous breaking of rotational symmetry. In these system, the
ground state exhibits a finite magnetization and a long-range order of
staggered magnetization, respectively. Thus, the ground state does not
exhibit the full rotational symmetry, but is only invariant under
rotations around the axis of magnetization. This is the spontaneous
breaking of rotational symmetry down to axial symmetry.  In the case
of the ferromagnet, ground states with different orientations of the
magnetization are inequivalent. The effective field theory for magnets
has been derived by Leutwyler~\cite{Leut93}.

In systems with a finite number of degrees of freedom, such as atomic
nuclei, spontaneous symmetry breaking does not occur in a strict sense
since arbitrarily small perturbations do not lead to a deformation of
the ground state. However, for nuclei in the rare earth region,
perturbations of the size of a few tens of keV can mix states with
different angular momenta. This is a particularly low energy scale
compared to other excitations, and it is two orders of magnitude
smaller than the nucleon separation energy.  In a technical sense,
spontaneous symmetry breaking does not take place because differently
oriented ground states of intrinsically deformed nuclei are unitary
equivalent to each other. A superposition (i.e. the projection) of
these states creates states with good angular momentum that are
invariant under the full rotation group~\cite{Ui1983}. This phenomenon
is well established in mean field calculations (see, e.g.,  
Refs.~\cite{Reinhard,Nazarewicz,Frauendorf}). However,
the small energy scale necessary to induce a symmetry breaking
justifies and motivates us to apply the ideas of spontaneous symmetry
breaking to this case.

For axially deformed nuclei, the rotational symmetry is spontaneously
broken. The ground state $\psi^{(0)}$ is only invariant under operations of the
subgroup ${\cal H}=SO(2)$, i.e.  
\be 
h\psi^{(0)} = \psi^{(0)} \quad\mbox{for $h\in {\cal H}$} \ , 
\ee 
but it is not invariant under general operations of the full rotation
group ${\cal G}=SO(3)$ of the Hamiltonian. Any rotation $g\in{\cal G}$
can be decomposed as $g=\tilde g h$ with $\tilde g\in {\cal G}$ and
$h\in {\cal H}$.  Two rotations $g=\tilde g h$ and $g'= \tilde g h'$
with $h, h'\in {\cal H}$ that differ from each other by an element of
the subgroup ${\cal H}$ yield the same state when applied to the
ground state $\psi^{(0)}$. Thus, such group elements must be
identified, and they form an equivalence class. This equivalence class
is the coset ${\cal G}/{\cal H}$. The coset $SO(3)/SO(2)$ can thus be
used to describe the low-energy degrees of freedom which change the
orientation of the ground state $\psi^{(0)}$. The corresponding
degrees of freedom are Nambu-Goldstone modes. In infinite systems,
these modes have zero mass and thus zero energy in the limit of
vanishing momenta. We will see below that the Nambu-Goldstone modes of
deformed nuclei generate rotational bands upon quantization.

The Nambu-Goldstone modes parameterize the coset $SO(3)/SO(2)$, and we
need to work out the basic expressions from which rotationally
invariant Lagrangians can be constructed.  This problem was solved for
the spontaneous breaking of chiral symmetry by
Weinberg~\cite{Weinberg}, and for general Lie groups by Coleman, Wess
and Zumino~\cite{CWZ}, and by Callan, Coleman, Wess and
Zumino~\cite{CCWZ}. For detailed reviews of this matter, the reader is
refered to Refs.~\cite{Weinberg2,Para,Miransky,Brauner}. In this
Section, we closely follow~\cite{Weinberg2}.

Let us parameterize a rotation $r\in SO(3)$ by the three
Euler angles $\alpha, \beta$, and $\gamma$ as~\cite{bible}
\be
\label{gEuler}
r(\alpha,\beta,\gamma) = e^{-i\alpha \hat{J}_z}e^{-i\beta \hat{J}_y}e^{-i\gamma \hat{J}_z} \ .
\ee
Here, $\hat{J}_k$, $k=x,y,z$ denote the appropriate components of the
angular momentum operator (i.e. the generators of the Lie group $SO(3)$). 
These operators fulfill the usual commutation relations
\be
[\hat{J}_j,\hat{J}_k] = i \sum_l \varepsilon_{jkl} \hat{J}_l \ . 
\ee
Let us assume that the subgroup ${\cal H}=SO(2)$ consists of the
rotations $h(\gamma)=\exp{(-i\gamma \hat{J}_z)}$, i.e. its generator
is $\hat{J}_z$.  This implies that the ground state $\psi^{(0)}$ has a
finite expectation value in the $z$-direction. Thus, any two rotations
$r(\alpha,\beta,\gamma)$ that differ by the Euler angle $\gamma$ from
each other yield the same state when acting on the ground state. The
coset $SO(3)/SO(2)$ thus consists of the rotations
\be
\label{gCoset}
g(\alpha,\beta) = e^{-i\alpha \hat{J}_z}e^{-i\beta \hat{J}_y}\ ,
\ee
and the Euler angles $\alpha$ and $\beta$ are the degrees of freedom of the 
Nambu-Goldstone modes.

The dynamics of the Nambu-Goldstone modes is determined by the time derivative
$\partial_t g(\alpha,\beta)$.  It is simpler to compute
\ba
\label{ginvdg}
g^{-1}(\alpha,\beta)\partial_t g(\alpha,\beta) &\equiv& 
i E_x \hat{J}_x +i E_y \hat{J}_y + i E_z \hat{J}_z \ ,    
\ea
as this is an element of the Lie algebra of the group ${\cal
  G}=SO(3)$. This defines the functions $E_k$, $k=x,y,z$. 

We want to construct Lagrangians that are invariant under rotations
and therefore need to study the transformation properties of the
functions $E_x$, $E_y$, and $E_z$. Let us act with a rotation
$r\in{\cal G}$ onto a rotation $g\in {\cal G}/{\cal H}$. This yields $rg$ 
which again can be decomposed into a product of two rotations $\tilde
g\in {\cal G}/{\cal H}$ and $h\in {\cal H}$
\be
\label{rg}
 r g = \tilde g(g,r) h(g,r) \ .
\ee 
Solving for $\tilde g$ yields 
\be
\tilde g(g,r) = r g h^{-1}(g,r) \ . 
\ee 
The rotations $\tilde g$ and $h$ are nonlinear (and complicated) functions
of the three Euler angles that define $r$ and the two Nambu-Goldstone modes that
define $g$. We have 
\ba
g^{-1}\partial_t g &=& (rg)^{-1}\partial_t (rg)\nonumber\\
&=& h^{-1} (\tilde g\partial_t \tilde g) h + h^{-1} \partial_t h \ .
\ea
In this derivation we used that $r$ is time independent, and we employed 
Eq.~(\ref{rg}). We solve for $(\tilde g\partial_t \tilde g)$ and find
\be
\label{trafo}
\tilde{g}^{-1}\partial_t \tilde g = h (g^{-1}\partial_t g) h^{-1} - (\partial_t h) h^{-1} \ . 
\ee
Similar to Eq.~(\ref{ginvdg}) we again have 
\be
\label{tildeginvdg}
\tilde{g}^{-1}\partial_t \tilde g \equiv i \tilde E_x \hat{J}_x +  i \tilde E_y \hat{J}_y 
+  i \tilde E_z \hat{J}_z \ , 
\ee
as this is an element of the Lie algebra of $SO(3)$.  We employ
Eq.~(\ref{tildeginvdg}) and Eq.~(\ref{ginvdg}) on the left and
right-hand side of Eq.~(\ref{trafo}), respectively, and observe that
$h\hat{J}_{x,y} h^{-1}$ is a linear combination of $J_{x,y}$ while
$h\hat{J}_z h^{-1}=\hat{J}_z$.  Note that the term $(\partial_t h)
h^{-1}$ on the right-hand side of Eq.~(\ref{trafo}) is also proportional to $\hat{J}_z$. 
Thus,
\ba
\label{ExEytrafo}
\tilde{E}_x\hat{J}_x +\tilde{E}_y\hat{J}_y 
&=& E_x \,h\hat{J}_x h^{-1} +E_y \,h\hat{J}_y h^{-1} \ , \\
\label{Eztrafo}
\tilde{E}_z\hat{J}_z &=& E_z \hat{J}_z  -i(\partial_t h) h^{-1} \ .
\ea
These equations show that under a general rotation $r$, the functions
$E_x$ and $E_y$ transform as the $x$ and $y$ components of a vector
under the rotation $h$ around the $z$-axis, while $E_z$ transforms as
a gauge field.  Indeed,
\be
h(\partial_t -iE_z\hat{J}_z)h^{-1} = \partial_t -i\tilde{E}_z\hat{J}_z \ .
\ee

Let us further illuminate these derivations. We express the rotation
$h$ as $h = \exp{(-i \gamma(g,r)\hat{J}_z)}$. Here the angle
$\gamma(g,r)$ is a (complicated) function of the Nambu-Goldstone
modes $\alpha$ and $\beta$ of the rotation $g$ and the three angles
that parameterize the rotation $r$.  We employ the
transformation~(\ref{ExEytrafo}) and find
\ba
\label{exey-trans}
\left(\begin{array}{c}
\tilde{E}_x \\
\tilde{E}_y\end{array}\right)
&=&  
\left(\begin{array}{cc}
\cos\gamma & \sin\gamma\\
-\sin\gamma & \cos\gamma\end{array}\right)
\left(\begin{array}{c}
E_x \\
E_y\end{array}\right) \ ,
\ea
while Eq.~(\ref{Eztrafo}) yields
\be
\label{EzExample}
\tilde{E}_z = E_z-\dot\gamma \ .
\ee
Note that Eq.~(\ref{exey-trans}) implies that
\ba
E_{\pm 1}\equiv E_x\mp iE_y
\ea
transforms as the upper and lower components of a vector (a spherical 
tensor of degree one), i.e. 
\ba
e^{i\gamma \hat{J}_z} E_{\pm 1} = e^{\pm i\gamma} E_{\pm 1} \ .
\ea
 
A Lagrangian consisting of combinations of the functions $E_x$, and
$E_y$ that are formally invariant under the subgroup ${\cal H}=SO(2)$
will thus be invariant under general rotations $r\in{\cal
  G}=SO(3)$~\cite{Weinberg2}. This is all we need for the construction
of Lagrangians involving the Nambu-Goldstone modes.
Equation~(\ref{EzExample}) shows that the function $E_z$ is not
invariant under a rotation, but merely changes by a total derivative.
The function $E_z$ will be important in cases where time reversal
invariance is broken by the ground state (i.e. for nuclei with a
finite ground-state spin).

Let us also consider the presence of a field $\psi$ that describes
physics at a higher energy scale.  The key is~\cite{Weinberg2} to rewrite
\ba
\label{so2_dphonon}
\psi  \equiv g(\alpha,\beta) \phi \ .
\ea
Here, $g$ is the element~(\ref{gCoset}) of the coset ${\cal G}/{\cal
  H}=SO(3)/SO(2)$, and $\phi$ is defined in terms of $\psi$ and $g$.
Due to the particular choice~(\ref{so2_dphonon}), a rotation $r\in
{\cal G}$ transforms $\psi\to \tilde \psi$ as
\be
\label{psi_trafo}
\tilde \psi \equiv r\psi = rg\phi = \tilde g h \phi \ . 
\ee
Here, we used Eq.~(\ref{rg}).  By definition (compare to
Eq.~(\ref{so2_dphonon})), we must also have $\tilde \psi \equiv \tilde
g \tilde \phi$. The comparison with Eq.~(\ref{psi_trafo}) shows that
under a rotation, the field $\phi$ transforms as $\phi\to\tilde\phi$
with
\be
\tilde \phi \equiv h\phi \ .
\ee
The time derivative of the field
\be
\partial_t\tilde \phi = (\partial_t h) \phi + h\partial_t \phi  
\ee
thus transforms as a gauge field. Employing Eq.~(\ref{Eztrafo}) we have
\be
h\left(\partial_t -iE_z\hat{J}_z\right)\phi 
= \left(\partial_t -i\tilde{E}_z\hat{J}_z\right)\tilde\phi \ .
\ee
Thus, the covariant derivative
\be
\label{covderiv}
D_t\equiv \partial_t -i E_z J_z \ ,
\ee 
when acting onto $\phi$, transforms properly under rotations and can
be employed in the construction of rotationally invariant Lagrangians
for the Nambu-Goldstone modes and the field $\psi$.  A Lagrangian
consisting of of $E_x$, $E_y$, $\phi$, and $D_t\phi$ that is formally
invariant under rotations of the subgroup ${\cal H}=SO(2)$ will be
invariant under the full action of the group ${\cal G}=SO(3)$. We now
recognize the advantage of employing a nonlinear realization of the
rotational symmetry. While the derivation of the basic building blocks
$E_x$, $E_y$, $\phi$, and $D_t\phi$ and their transformation
properties are somewhat more complicated than for the well-known
linear representations, the construction of rotationally invariant
Lagrangians can be achieved by constructing Lagrangians that (at first
sight) only appear to exhibit axial symmetry. This will make the
resulting effective theory computationally tractable. This is in
contrast to algebraic models which are linear representations of the
rotational symmetry and employ dynamical symmetries to remain
computationally feasible~\cite{Rowe}.

All that remains is to actually compute the functions $E_k$,
$k=x,y,z$.  We use the Baker-Campbell-Hausdorff formula and compute
the left-hand-side of Eq.~(\ref{ginvdg}) directly from the
expression~(\ref{gCoset}). This yields
\ba
\label{Exyzfun}
E_x &=& \dot\alpha \sin\beta \ ,\nonumber\\
E_y &=& -\dot\beta \ ,\nonumber\\
E_z &=& -\dot\alpha \cos\beta \ .
\ea
From a practical point of view, these expressions are the main result
of this Section. Note that the components $E_k$ can be viewed as angular
velocities.

The invariance under rotation implies that angular momentum is a
conserved quantity. To become more familiar with the nonlinear
realization of rotational symmetry, we compute this conserved quantity via
Noether's theorem.  Under infinitesimal rotations around the $k$ axis
($k=x, y, z$) by an angle $\delta \omega_k$, the Euler angles $\alpha$
and $\beta$ change by the infinitesimal amounts $\delta\alpha$, and
$\delta\beta$, respectively.  One finds (see~\ref{app:trafo} for details)

\ba
\label{trafo_euler}
\left(\begin{array}{c}
\delta\alpha \\
\delta\beta
\end{array}\right)
= \hat{M}
\left(\begin{array}{c}
\delta \omega_x \\
\delta \omega_y \\
\delta \omega_z
\end{array}\right) \ , 
\ea
with
\ba
\hat{M}=
\left(\begin{array}{ccc}
-\cot\beta \cos\alpha & -\cot\beta \sin\alpha & 1 \\
-\sin\alpha & \cos\alpha & 0
\end{array}\right) \ .
\ea

The Lagrangian $L(\dot\alpha,\dot\beta,\beta)$ of the Nambu-Goldstone modes is invariant 
under rotations. We apply Noether's theorem and find the conserved
quantities (see~\ref{app:noether} for details)
\ba
\left(\begin{array}{c}
Q_x \\
Q_y \\
Q_z
\end{array}\right) = \hat{M}^T 
\left(\begin{array}{c} p_\alpha \\ p_\beta\end{array}\right) 
\ .
\ea  
Here, 
\ba
p_\alpha &=& {\partial L\over\partial\dot\alpha} \ , \\
p_\beta &=& {\partial L\over\partial\dot\beta} 
\ea
are the momenta conjugate to $\alpha$ and $\beta$, respectively.  Thus, we have
\ba
\label{conserved_NG}
Q_x &=& -p_\beta \sin\alpha -p_\alpha \cot\beta \cos\alpha \ , \nonumber\\
Q_y &=& p_\beta \cos\alpha -p_\alpha \cot\beta \sin\alpha \ , \nonumber\\
Q_z &=& p_\alpha \ ,
\ea  
and the total angular momentum squared is
\ba
\label{angmom}
 Q^2 \equiv Q_x^2+Q_y^2+Q_z^2
= p_\beta^2 +{p_\alpha^2\over \sin^2\beta} \ . 
\ea
Clearly, $Q^2$ is the squared angular momentum of a rotor, and we will see 
in Sect.~\ref{sec:NG} that the leading order Hamiltonian of the 
Nambu-Goldstone modes is proportional to this quantity.

In what follows, we will continue to employ the Euler angles $\alpha$
and $\beta$ as the Nambu-Goldstone modes.  This is a convenient but arbitrary
choice. The algebraic transformation laws derived in this subsection
are independent of this particular choice as it only depends on the
pattern of the symmetry breaking $SO(3)\to SO(2)$. It is only the
explicit expressions~(\ref{Exyzfun}) for the functions $E_{x}$,
$E_{y}$, and $E_{z}$, respectively, that depend on this
parameterization of the coset.

Let us briefly contrast the effective theory based on nonlinear
realizations with the earlier phenomenological approaches.  The Bohr
Hamiltonian~\cite{Bohr} and its generalizations~\cite{Greiner} model
the collective states of atomic nuclei by surface vibrations. In
leading order, these are quadrupole phonons. Within this approach, one
transforms from the laboratory coordinate system to the co-rotating
(or body fixed) coordinate system.  Thereby one introduces three Euler
angles and the two moments of inertia that identify an axially
deformed nucleus as the relevant degrees of freedom.  Within the
effective theory presented in this paper, different variables are
chosen. In the case of axially deformed nuclei, only two Euler angles
are relevant for the description of low-energy excitations. As we will
see, the nonlinear realization naturally corresponds to the
spontaneously broken rotational symmetry and facilitates the
introduction of a power counting.

\section{Nambu-Goldstone modes}
\label{sec:NG}

\subsection{Even-even nuclei}
\label{ng_ee_nuclei}

Let us assume that the ground state is invariant under time reversal
(as is the case for even-even nuclei). The simplest Lagrangian is
second order in the time derivative, and below we will see that this
Lagrangian is indeed of leading order.  In leading order (LO), the
Lagrangian of the Nambu-Goldstone modes thus is
\ba
\label{nlag}
L_{\rm LO}^{(ee)} ={C_0\over 2}\left(E_x^2 + E_y^2\right) 
= {C_0\over 2}\left(\dot{\beta}^2 +\dot{\alpha}^2\sin^2{\beta}\right) \ . 
\ea
Here $C_0$ is a low-energy
constant to be determined by fit to data. A Legendre transformation 
yields the
Hamiltonian
\ba
\label{ham_cl}
H={p_\beta^2\over 2C_0} +{p_\alpha^2\over 2 C_0\sin^2\beta} \ . 
\ea
Here, we employed the canonical momenta $p_\beta = \partial
L/\partial\dot\beta$ and $p_\alpha = \partial L/\partial\dot\alpha$.
The Hamiltonian~(\ref{ham_cl}) is obviously proportional to the
squared angular momentum~(\ref{angmom}).  The quantization in
curvilinear coordinates is well known~\cite{Podolsky}, see~\ref{app:legendre} for details. 
This yields the Hamiltonian (we use units where $\hbar=1$)
\ba 
\label{nham}
\hat{H}&=& -{1\over 2C_0} \left(\partial^2_\beta + \cot{\beta}\partial_\beta 
+{1\over \sin^2\beta}\partial^2_\alpha  
\right)\ .
\ea
Comparison with the classical Hamiltonian~(\ref{ham_cl}) thus yields
\ba
\label{momenta}
p_\beta^2 &=& -{1\over\sin\beta} \partial_\theta \sin{\beta} \partial_\beta \ , 
\nonumber\\
p_\alpha &=& -i\partial_\alpha \ .
\ea
The eigenfunctions of the Hamiltonian~(\ref{nham}) 
are spherical harmonics, i.e. 
\be
\hat{H} Y_{lm}(\beta,\alpha)={l(l+1)\over 2C_0} Y_{lm}(\beta,\alpha) \ , 
\ee
and $l=0,1,2,\ldots$ thus yields a rotational spectrum.  Note that the
low-energy constant $C_0$ is, within the collective model, 
associated with the moment of inertia.

For even-even nuclei, only even values of $l$ are permitted for the
ground-state band. This can be understood as follows.  The ground
state of even-even nuclei is not only invariant under rotations
$h\in{\cal H}=SO(2)$ but also under discrete operations such as
parity, or a rotation about $\pi$ around an axis perpendicular to
axial symmetry axis.  Thus, we must augment the invariant subgroup
${\cal H}$ by these discrete operations, and this modifies the coset
accordingly.  As a result, the angles $(\beta,\alpha)$ and
$(\pi-\beta,\pi+\alpha)$ have to be identified, and this limits the
values of $l$ to even numbers~\cite{BM}. 

\subsection{Odd-mass and odd-odd nuclei}
\label{subsub:odd}   
Odd nuclei have a half-integer ground-state spin, while odd-odd nuclei
can also exhibit a nonzero integer ground-state spin. Thus, the ground
state is not invariant under time reversal. This modifies the
low-energy effective Lagrangian as terms that are not invariant under
time reversal need to be considered. Such terms consist of only one
time derivative, and we need to consider the
functions~(\ref{Exyzfun}).  The transformation
properties~(\ref{ExEytrafo}) and (\ref{Eztrafo}) show that the
functions $E_x$, $E_y$, and $E_z$ are not invariant under rotations.
However, the function $E_z$ only changes by a total derivative, see
Eq.~(\ref{EzExample}). Thus, the action changes by an irrelevant
phase. In quantum field theory, such a function is known as a
Wess-Zumino term.  In our case, the Wess-Zumino term $L_{\rm WZ}$ is
\ba
\label{WZ}
L_{\rm WZ} \equiv qE_z = -q \dot{\alpha}\cos{\beta} \ .
\ea
Recall that a low-energy Lagrangian can be understood as resulting
from integrating out high-energy fermion modes in a more fundamental
Lagrangian~\cite{DBKaplan}. In the case that the considered fermion
system consists of an odd number of fermions, or has a finite
ground-state spin, the resulting low-energy Lagrangian must reflect
this behavior. Thus, the appearance of a corresponding
symmetry-breaking term is unavoidable.  For details, the reader is
refered to Refs.~\cite{Stone,Aitchinson}.  In
Sect.~\ref{sec:fermions}, we will couple fermions to the
Nambu-Goldstone modes and see that the relevant term is indeed
proportional to $E_z$, as stated in Eq.~(\ref{WZ}).

Let us consider the transformation properties of the Wess-Zumino term.
Under a rotation by $\delta\omega_k$ around the $k$ axis ($k=x,y,z$),
the Wess-Zumino Lagrangian $L_{\rm WZ}$ changes by (see~\ref{app:noether} for details)
\be 
\label{delta_LWZ}
\delta L_{\rm WZ} = 
q \left( \delta \omega_x \partial_t\left({\cos\alpha\over\sin\beta}\right) 
+ \delta \omega_y \partial_t\left({\sin\alpha\over\sin\beta}\right) \right) \ ,
\ee
and this is obviously a total derivative. The application of
Noether's theorem yields the conserved quantities
\ba
\label{conserved_WZ}
Q_x &=& -{\cos\alpha\over\sin\beta}q
-p_\beta \sin\alpha -p_\alpha \cot\beta \cos\alpha \ , \nonumber\\
Q_y &=& -{\sin\alpha\over\sin\beta}q
+p_\beta \cos\alpha -p_\alpha \cot\beta \sin\alpha \ , \nonumber\\
Q_z &=& p_\alpha \ ,
\ea  
which are the components of the angular momentum. 
The total angular momentum squared is 
\ba
\label{angmom_WZ}
Q^2 \equiv Q_x^2 + Q_y^2 +Q_z^2
= p_\beta^2 + {1\over\sin^2\beta}\left(p_\alpha +q\cos\beta\right)^2 
+q^2 \ .
\ea

With the Wess-Zumino term added, the leading order Lagrangian becomes
\ba
\label{nlag_univ}
L_{\rm LO}= L_{\rm LO}^{(ee)} + L_{\rm WZ}
={C_0\over 2}\left(\dot{\beta}^2 +\dot{\alpha}^2\sin^2{\beta}\right)  
- q \dot{\alpha}\cos{\beta} \ . 
\ea
The Legendre transformation yields the classical Hamiltonian
\be
\label{ham_LO_class}
H_{\rm LO}={p_\beta^2\over 2C_0} 
+{\left(p_\alpha+q\cos{\beta}\right)^2 \over 2 C_0\sin^2\beta} \ .
\ee
The comparison with the angular momentum~(\ref{angmom_WZ}) shows that 
$H_{\rm LO}=(Q^2-q^2)/(2C_0)$. 
We employ the quantization~(\ref{momenta}), and obtain
\ba
\label{ham_LO}
\hat{H}_{\rm LO}=-{1\over 2C_0\sin\beta}\partial_\beta\sin\beta \partial_\beta 
+{1\over 2C_0\sin^2\beta}\left(-i\partial_\alpha+q\cos{\beta}\right)^2 \ .
\ea
The eigenfunctions of this Hamiltonian are products $e^{-i\alpha
  m}d^l_{mq}(\beta)$. Here $d^l_{mq}$ denotes the ``little'' Wigner $d$
function~\cite{bible}, i.e. the Wigner $D$ function is defined as
$D^l_{mq}(\alpha,\beta,\gamma)\equiv e^{-i m \alpha} d^l_{mq}(\beta)
e^{-i q \gamma}$. Details are presented in the~\ref{app:dgl}. Thus,
\ba
\hat{H}_{\rm LO}\, d^l_{mq}(\beta) e^{-i\alpha m} 
&=& E_{\rm LO} (q,l) d^l_{mq}(\beta) e^{-i\alpha m} \ ,
\ea
and the eigenvalues are
\ba
\label{ergLO}
E_{\rm LO}(q,l)&=&{l(l+1)-q^2\over 2C_0} \nonumber\\
&=& {|q|\over 2 C_0} + {l(l+1)-|q|(|q|+1)\over 2 C_0} \ .
\ea    
Here, $q\ne 0$ must be integer or half integer, while $l$
formally assumes the values $l=|q|, |q|+1, |q|+2,\ldots$, and $m=-l,
-l+1,\ldots l$.  (For $q=0$, $l$ only assumes even values.) 

Apart from the irrelevant constant $|q|/(2C_0)$, the comparison of
Eq.~(\ref{ergLO}) with Eq.~(\ref{master}) shows that we have to
identify the low-energy constant $q$ with the ground-state spin $I_0$
of the nucleus under consideration, i.e.  $|q|=I_0$. Thus, the spin
$I_0$ of the ground-state and the spacing between the lowest two
states in the rotational spectrum fix the low-energy constants $q$ and
$C_0$.  This unified description of odd and even nuclei within an
effective theory is very encouraging.  The effective
Hamiltonian~(\ref{ham_LO}) yields the correct low-energy description
of axially deformed nuclei.  The case of even-even nuclei (and odd-odd
nuclei with zero ground-state spin) is particularly simple as
$q=I_0=0$, while odd-mass nuclei or odd-odd nuclei with nonzero
ground-state spin $I_0$ have a more complicated Hamiltonian due to
$q=I_0 > 0$.

\subsection{Next-to-leading order}
\label{sub2_nlo}
We need to establish a power counting. Let us denote the low-energy scale
associated with the Nambu-Goldstone modes as $\xi$. Thus, the leading-order
energy~(\ref{ergLO}) scales as $E_{\rm LO} \sim \xi$, and we get the
following estimates
\ba
\label{scale_NG}
C_0&\sim& \xi^{-1} \ ,\nonumber\\
p_\beta \sim p_\alpha \sim q &\sim& \xi^0 \ , \nonumber\\
\dot\beta \sim \dot\alpha \sim E_{x,y,z}&\sim& \xi \ . 
\ea
The estimate in the second line of Eq~(\ref{scale_NG}) is due to the
quantization of angular momentum (recall that $\hbar=1$), and the
estimate in the third line of Eq.~(\ref{scale_NG}) follows from
the usual relationship between velocities and momenta. It is clear
that the Wess-Zumino term scales as $L_{\rm WZ}\sim \xi$, and is
indeed of leading order.

The Nambu-Goldstone modes of energy $\sim \xi$ are well separated from
the high energy (breakdown) scale $\Omega\gg \xi$ which is associated
with degrees of freedom (vibrational, pairing or single-particle
degrees of freedom) we have omitted from our theory. In an effective
theory, these omitted terms manifest themselves through interactions
between the Nambu-Goldstone modes.  At next-to-leading order (NLO),
higher derivatives of the Nambu-Goldstone modes
appear~\cite{Weinberg2,Leut93}, and one new term enters the Lagrangian
\ba
\label{nlolag}
L_{\rm NLO}&=&L_{\rm LO} + {C_2\over 4}\left(E_x^2+E_y^2\right)^2  \ .
\ea
Here, $C_2$ is the corresponding low-energy constant. The ratio
$C_2/C_0$ has units of energy$^{-2}$, and is assumed to be due to
omitted physics at the breakdown scale. Thus, the dimensional analysis
yields
\be
\label{scale_c2}
{C_2\over C_0}\sim \Omega^{-2} \ , 
\ee
and 
\be
\label{hierarchy}
{C_2\over C_0}\left(E_x^2+E_y^2\right) \sim\left({\xi\over\Omega}\right)^2 \ll 1 \ . 
\ee
The last equation indicates that the next-to-leading-order correction is suppressed by
two powers of $\xi/\Omega$ compared to the leading-order terms.

For the computation of the Hamiltonian at next-to-leading order, we 
employ the conjugate momenta
\ba
p_\beta &\equiv& {\partial L_{\rm NLO}\over \partial\dot\beta} 
= \left(C_0+C_2 \left(E_x^2+E_y^2\right)\right)\dot\beta   \ ,\\
p_\alpha &\equiv& {\partial L_{\rm NLO}\over \partial\dot\alpha}
=\left(C_0+C_2 \left(E_x^2+E_y^2\right)\right) \dot\alpha\sin^2\beta 
-q\cos{\beta}  \ ,
\ea
and apply Eq.~(\ref{hierarchy}) consistently in the Legendre transform.  
This yields the Hamiltonian at next-to-leading order
\ba
H_{\rm NLO}= \left(1-{C_2\over C_0^2} H_{\rm LO}\right) H_{\rm LO}   \ .
\ea   
Here, $H_{\rm LO}$ is the leading-order Hamiltonian~(\ref{ham_LO_class}).
Thus, the spectrum of the Nambu-Goldstone modes becomes
\ba
E_{\rm NLO} &=& \left(1-{C_2\over C_0^2} E_{\rm LO}\right) E_{\rm LO}   \ ,
\ea 
where $E_{\rm LO}$ is given in Eq.~(\ref{ergLO}).  Thus, the 
spectrum at next-to-leading order 
is a polynomial of degree two in the right-hand side of
Eq.~(\ref{master}).  This is exactly as given by Bohr and
Mottelson~\cite{BM}. The low-energy constant $C_2$ can be determined by fitting the
spacing between the rotational state with $l=I_0+4$ and $l=I_0$.  For
deformed rare earth nuclei, the ratio $\xi/\Omega\approx 10^{-1}$,
and this explains the high quality of many rotational bands, i.e. the
next-to-leading-order contribution is a small correction for angular momenta $l$ with
$l\ll\Omega/\xi$.  Higher corrections at next-to-next-to-leading order include terms
of the form
\be
C_4\left(E_x^2 +E_y^2\right)^3 \ , 
\ee
and the dimensional analysis yields $C_4/C_2\sim \Omega^{-2}$. Thus,
another factor $(\xi/\Omega)^2\approx 10^{-2}$ separates these
contributions from the contributions at next-to-leading order. However
for rotational states with high angular momentum $l\approx {\cal
  O}(\Omega/\xi)$, the description in terms of the Nambu-Goldstone
modes must break down, as higher order terms become large. In this
case, one needs to include degrees of freedom hat are higher in energy
such as vibrations, pairing effects, and nucleonic excitations. 
This is the subject of the next two sections.

Note finally that the results presented in this section can also be
obtained by simpler means. An alternative derivation is given in~\ref{app:ng}.

\section{Coupling of quadrupole bosons to the Nambu-Goldstone modes}
\label{sec:vib}

In this Section, we couple higher-energetic phonon degrees of
freedom to the Nambu-Goldstone modes. The appropriate phonons are quadrupole
vibrations, as evident from the spins and parities of low-energy states. 
This results from the observation that spectra of
even-even nuclei exhibit $I_0=2$ band heads at the energy scale
$\Omega$. In this section we couple the quadrupole phonons (whose ground state
spontaneously breaks the rotational symmetry) to the Nambu-Goldstone modes.

\subsection{Quadrupole phonons}
We consider a quadrupole field $\psi$ (spin-two boson) with complex
components $\psi_\mu$, $\mu=-2,\ldots,2$. Invariance under time
reversal implies
\be
\psi_{-\mu}^* = (-1)^\mu\psi_\mu \ ,  
\ee
such that we deal with five degrees of freedom. 

As discussed in Section~\ref{sec:coset}, we parameterize the phonon
as $\psi=g\phi$ with $g\in{\cal G}/{\cal H}=SO(3)/SO(2)$.
By assumption, the ground state $\phi^{(0)}$ spontaneously breaks
rotational symmetry, but remains invariant under the operations of the
subgroup ${\cal H}=SO(2)$ of the rotation group $SO(3)$. Again, we choose
$\hat{J}_z$ as the generator of the subgroup ${\cal H}$.
Consequently, the ground-state expectation value of the field $\phi$
fulfills
\be
\langle \phi^{(0)} \rangle_\mu =v\delta_\mu^0 \ , 
\ee
with $v\ne 0$, and is obviously invariant under rotations around the
(arbitrarily chosen) $z$ axis.  

We have to account for the fact that the Nambu-Goldstone modes result
from the symmetry breaking of the quadrupole phonon $\phi$.  Thus, we
parameterize the five components of $\psi=g\phi$ in terms of the two
Nambu-Goldstone modes $\alpha$ and $\beta$, and three
non-Nambu-Goldstone modes as
\ba
\label{phi}
\phi=\left(\begin{array}{c} \phi_2\\ 0\\ \phi_0\\ 0\\ \phi_2^*\end{array}\right) \ .
\ea
The three non-Nambu-Goldstone modes are parameterized by a complex
component $\phi_2$ and the real component $\phi_0$. The
choice~(\ref{phi}) for the non-Nambu-Goldstone mode is appropriate~\cite{Weinberg2} 
as $\phi$ is
``independent'' of the Nambu-Goldstone modes, i.e.  it is orthogonal
to an infinitesimal rotation of the ground state induced by the
generators $\hat{J}_k$, $k=x,y$ that do not belong to the Lie algebra
of the subgroup ${\cal H}=SO(2)$
\ba
\phi^\dagger (\hat{J}_{k} \langle \phi^{(0)}\rangle) = 0 
\quad\mbox{with $k=x,y$.}
\ea 

The ``building blocks'' for rotationally invariant Lagrangians are
$D_t \varphi$, $D_t\phi_2$, $D_t\phi_{-2}$, $E_1$, $E_{-1}$, $\varphi_0$,
$\phi_2$, and $\phi_{-2}=\phi_2^*$. Any Lagrangian in these
quantities that is formally invariant under $SO(2)$ will indeed
exhibit full rotational invariance because of the nonlinear
realization of the symmetry.

\subsection{Transformation properties}

Recall the transformation properties and the conserved quantities
of our effective theory. Under a rotation $r$, the field $\phi$ transforms as
\ba
\phi\to h(\gamma)\phi = e^{-i\gamma \hat{J}_z} \phi \ .
\ea 
Here $\gamma =\gamma(\alpha,\beta,r)$ is a complicated angle of the
rotation $r$ and the Nambu-Goldstone modes. Note that the component
$\phi_0$ is invariant under rotations. For infinitesimal rotations by
angles $\delta\omega_k$ around the $k$-axis, we have (see~\ref{app:trafo} for details)
\be
\gamma = {\cos\alpha\over\sin\beta} \delta\omega_x 
+{\sin\alpha\over\sin\beta} \delta\omega_y \ .
\ee
Note that $\gamma=0$ for a rotation around the $z$-axis. We formally apply 
Noether's theorem to the Lagrangian $L(\dot\alpha,\dot\beta,\beta,\dot\phi,\phi)$ 
and find the conserved quantities 
\ba
\label{conserved_phi}
Q_x &=& -2{\cos\alpha\over\sin\beta}l_2
-p_\beta \sin\alpha -p_\alpha \cot\beta \cos\alpha \ , \nonumber\\
Q_y &=& -2{\sin\alpha\over\sin\beta}l_2
+p_\beta \cos\alpha -p_\alpha \cot\beta \sin\alpha \ , \nonumber\\
Q_z &=& p_\alpha \ .
\ea  
Here, we decomposed $\phi_2 = \phi_{2r}+ i\phi_{2i}$ into its real and
imaginary part, respectively, denoted the momentum conjugate to
$\phi_{2r}$ ($\phi_{2i}$) by $p_{2r}$ ($p_{2i}$), and 
employed the angular momentum 
\be
\label{l2}
l_2\equiv \left(\phi_{2r}p_{2i}-\phi_{2i}p_{2r}\right) \ .  
\ee
The total angular momentum squared is
\ba
\label{q2}
Q^2 &=& p_\beta^2 +{1\over \sin^2\beta}\left(p_\alpha^2 
+4p_\alpha l_2\cos\beta+ 4l_2^2\right) \nonumber\\
&=& p_\beta^2 +{1\over \sin^2\beta}\left(p_\alpha^2 
+2l_2\cos\beta\right)^2 +(2l_2)^2 \ . 
\ea

\subsection{Power counting}
Let us consider the power counting. There are two possibly distinct
breakdown scales for our effective theory. First, the restoration of
rotational symmetry at high excitation energies signals the breakdown
of the effective theory. This scale results from the scales $\xi$ and
$\Omega$ that describe the quadrupole phonon $\psi=g\phi$. Second,
additional degrees of freedom enter at an energy scale
$\Lambda\gg\Omega$, and their effect also needs to be considered. (We
neglect, for instance, collective phonons of higher multipolarity, pairing, and
single-particle degrees of freedom.) We will focus here primarily on
the restoration of rotational symmetry.  The effects from not
included high-lying degrees of freedom will be neglected by formally
sending $\Lambda\to\infty$. 

The field~(\ref{phi}) spontaneously
breaks the rotational symmetry, and we separate the vacuum expectation 
value $v$ from the fluctuating contribution $\varphi_0$ as 
\be
\label{smallphi}
{\varphi_0}\equiv \phi_0-v \ . 
\ee
We have the following scaling relations
\ba
\label{scale_phi}
v   \sim \phi_0 &\sim& \xi^{-1/2} \ , \nonumber\\
\varphi_0 \sim \phi_2 &\sim& \Omega^{-1/2} \ , \nonumber\\ 
D_t\varphi_0\sim D_t\phi_2&\sim&\Omega^{1/2} \ , \nonumber\\
\dot{\varphi}_0= \dot{\phi}_0 \sim \dot{\phi}_2 &\sim& \Omega^{1/2} \ ,
\ea
in addition to the relations~(\ref{scale_NG}). 
The finite vacuum expectation value $v$ must clearly be associated
with the low-energy scale $\xi$, while the fluctuations of the field
$\phi$ are associated with the high-energy scale $\Omega$, and are
necessarily much smaller than the vacuum expectation value. Indeed, we
have $\varphi_0/v \sim \phi_2/v\sim(\xi/\Omega)^{1/2}\ll 1$. 

Let us briefly discuss the case of a finite breakdown scale
$\Lambda\gg\Omega$. Consider as an example the term $C\dot{\phi}_2^4$.
Here, $C$ has dimensions of energy$^{-1}$, and it must scale as
$C\sim\Lambda^{-1}$. Thus, the contribution $C\dot{\phi}_2^4\sim
\Omega^2/\Lambda$ is of next-to-leading order, i.e. it corrects the
leading-order term that scale as $\Omega$.  By sending
$\Lambda\to\infty$, we neglect such terms. However, it is clear that a
full-fledged effective theory needs to deal with them at some point in
the power counting.  The importance of such terms depends on how the
ratio $\xi/\Omega$ that governs the validity of the spontaneous
symmetry breaking compares to the ratio $\Omega/\Lambda$ that governs
the relevance of neglected degrees of freedom.

The construction of low-energy Lagrangians in the presence of a
spontaneously broken symmetry has been discussed in
Section~\ref{sec:coset}, with a focus on kinetic terms. We also
need to discuss the form of admissible potentials $V(\phi)$. A
rotationally invariant potential $V(\phi)$ consists of expressions
involving $\phi$ that are formally invariant under rotations around
the $z$ axis.  Furthermore, the potential must exhibit spontaneous
symmetry breaking.  We expand $V=V_{\rm LO}+V_{\rm NLO}+\ldots$. 
In leading order we have
\be
\label{potLO}
V_{\rm LO}(\phi) = {\omega_0^2\over 2}(\phi_0-v)^2 
+  {\omega_2^2\over 4}|\phi_2|^2 \ .
\ee
The low energy constants must scale as  $\omega_0\sim \omega_2\sim\Omega$ 
to yield the ground-state expectation value $\langle V_{\rm LO}\rangle\sim\Omega$.

For the construction of next-to-leading order potential terms, we need
to determine the breakdown scale for our effective theory. With
increasing energy, the fluctuations $\varphi_0$ grow in size, and the
effective theory breaks down once $\varphi_0\approx v$, since this
implies a restoration of the spherical symmetry.  Likewise, large
excitation energies correspond to a large amplitude $|\phi_2|\approx
v$, again resulting in a restoration of the spherical symmetry and in
the breakdown of the effective theory. The minimum kinetic energy of a
large-amplitude field $\phi$ is $v^{-2}\sim\xi$, while its potential
energy is $\langle V_{\rm LO}\rangle \sim \Omega^2/\xi \gg \Omega$. We will see that 
the power counting has to employ the kinetic energy scale.

Let us make a polynomial expansion of the potential
\be
\label{potential}
V=V_{\rm LO} + \sum_{k+2l>2} v_{kl}\varphi_0^k |\phi_2|^{2l} \ .
\ee
Here, $k, l$ are integers, and it is understood that only terms with
$k+2l>2$ are being summed over. (The leading order terms are $k+2l=2$.) 
Clearly, the potential exhibits
spontaneous symmetry breaking and is formally invariant under $SO(2)$.
Due to the nonlinear realization of the $SO(3)$ symmetry, it is also
invariant under $SO(3)$.

Note that $v_{kl}\varphi_0^k|\phi_2|^{2l}/V_{\rm LO}$ is dimensionless. This
implies that $v_{kl}/\Omega^2$ has dimension of energy$^{l-1+k/2}$. For
the power counting we have to assume that this energy scale has to be
identified with the kinetic energy scale.  Thus, we assume 
\be
v_{kl}\sim \Omega^2 \xi^{l-1+k/2} \ , 
\ee
and find
\be
v_{kl}\varphi_0^k|\phi_2|^{2l}\sim \Omega \left({\xi\over\Omega}\right)^{l-1+k/2} \ . 
\ee
This establishes the power counting for the potential.  Clearly, for
large amplitudes $\varphi_0,|\phi_2|\sim\xi^{-1/2}$ each potential
term of the sum~(\ref{potential}) is of order $\Omega$, signaling the
break down of the effective theory.

Our power counting is valid for well deformed nuclei that are
``rigid'' rotors , i.e. nuclei for which $\omega_0,\omega_2\gg \xi$.
Some deformed nuclei are not in this category. So-called $\gamma$-soft
nuclei, for instance, do not exhibit a separation of scale between
excitations within a rotational band (energy scale $\xi$) and the
vibration with frequency $\omega_2$. For these nuclei, the power
counting is different.

\subsection{Even-even nuclei}

Let us consider the case where the symmetry-breaking ground state is invariant under time reversal.
The following kinetic terms with two time derivatives are invariant
under rotations
\ba
(D_t\varphi_0)^2 &=& \dot{\varphi}_0^2 \nonumber\\
D_t\phi_2 D_t\phi_{-2} &=& \left|\dot{\phi}_2\right|^2
+4{\rm Im}\left(\dot{\phi}_2\phi_2^*\right)E_z
+4|\phi_2|^2E_z^2 \nonumber\\
E_x^2 +E_y^2 &=& \dot{\beta}^2+\dot{\alpha}^2\sin^2\beta \ .
\ea
Note that, in leading order, the covariant derivative is simply the 
usual time derivative. 
 
\subsubsection{Leading order}

In leading order (${\cal O}(\Omega)$), the Lagrangian is
\ba
\label{Llo}
L_{\rm LO}={1\over 2}\dot{\varphi_0}^2 + \left|\dot{\phi}_2\right|^2 
-{\omega_0^2\over 2}\varphi_0^2  - {\omega_2^2\over 4} |\phi_2|^2 \ .
\ea
This Lagrangian describes harmonic vibrations of the quadrupole
degrees of freedom. The Legendre transform yields the Hamiltonian
\be
\label{HLOvib}
H_{\rm LO} = {1\over 2}p_0^2 + {1\over 4}\left(p_{2r}^2 +p_{2i}^2\right)
+{\omega_0^2\over 2}\varphi_0^2  + {\omega_2^2\over 4} \left(\phi_{2r}^2 +\phi_{2i}^2\right) \ .
\ee 
Here, $\phi_{2r}$ and $\phi_{2i}$ are real variables that denote the 
real and imaginary parts of $\phi_2$, respectively, i. e. 
\ba
\phi_2=\phi_{2r}+i\phi_{2i} \ .
\ea
The momenta are defined as
\ba
p_{0}\equiv {\partial L_{\rm LO}\over \partial \dot{\varphi}_{0}} \ , \quad
p_{2r}\equiv {\partial L_{\rm LO}\over \partial \dot{\phi}_{2r}} \ , \quad
p_{2i}\equiv {\partial L_{\rm LO}\over \partial \dot{\phi}_{2i}} \ .
\ea
One might wonder whether the most general leading-order Lagrangian
should not have dimensionless low-energy constants in front of its
kinetic terms. Note, however, that any such constants could be
absorbed by a redefinition of the variables $\varphi_0$ and
$\phi_2$, and a rescaling of the oscillator frequencies $\omega_0$ and
$\omega_2$. 

The leading order part of the Hamiltonian~(\ref{HLOvib}) describes an
axially symmetric harmonic oscillator in three dimensions. The
canonical quantization rules $p_{0}\to -i\partial_{\varphi_0}$,
$p_{2r}\to -i\partial_{\phi_{2r}}$, and $p_{2i}\to
-i\partial_{\phi_{2i}}$ yields the Hamiltonian operator $\hat{H}_{\rm
  LO}$.  It is of advantage to seek not the Cartesian eigenstates but
rather eigenstates which reflect the axial symmetry. To his purpose,
we write $\phi_2 = \varphi_2 e^{i\gamma}$ and obtain the leading-order
Hamiltonian
\ba
\hat{H}_{\rm LO}&=& 
-{1\over 2}\partial_{\varphi_0}^2 + {\omega_0^2\over 2}\varphi_0^2 
+ {1\over 4}\left(
-\partial_{\varphi_2}^2 -{1\over\varphi_2}\partial_{\varphi_2}+ 
{\hat{l}_2^2\over \varphi_2^2}
+\omega_2^2\varphi_2^2 \right) \ .
\ea
Here, 
\be
\hat{l}_2 = -i\partial_\gamma 
\ee
is the operator corresponding to the classical expression~(\ref{l2}). 
The eigenstates and eigenenergies fulfill
\be
\hat{H}_{\rm LO}|n_0 n_2 l_2\rangle = E(n_0, n_2, l_2) |n_0 n_2 l_2\rangle \ .
\ee
The energies are
\be
\label{Evib}
E(n_0, n_2, l_2)= \omega_0\left(n_0+{1\over 2}\right) 
+{\omega_2\over 2}(2n_2 + |l_2| +1) \ . 
\ee
Here, $n_0=0,1,2,\ldots$ denotes the quantum number in the
$\varphi_0$-coordinate (i.e. excitation along the symmetry axis),
$n_2=0,1,2,\ldots$ denotes the radial quantum number associated with
the radius $\varphi_2=(\phi_{2r}^2 +\phi_{2i}^2)^{1/2}$, while $l_2=0,\pm 1,\pm
2,\ldots$ denotes azimuthal quantum number, i.e. the quantum number of the operator $\hat{l}_2$. 
Clearly, the
energies~(\ref{Evib}) are of order $\Omega$.

\subsubsection{Next-to-leading order}

In next-to-leading order (${\cal O}(\xi)$), the Lagrangian is
\ba
\label{Lnlo}
L_{\rm NLO}&=&L_{\rm LO} 
+  {v^2\over 2}\left(E_x^2+E_y^2\right) 
-4 E_z {\rm Im}\left(\dot{\phi}_2{\phi}_2^*\right) 
+ R_{\rm NLO}
\ea
with 
\ba
\label{Rnlo}
R_{\rm NLO}&=&
\left(Z_{00}\varphi_0^2 +Z_{02}|\phi_2|^2\right)\dot{\varphi}_2^2
+\left(Z_{20}\varphi_0^2 +Z_{22}|\phi_2|^2\right)|D_t{\phi}_2|^2\nonumber\\
&-&\sum_{k+2l=3,4} v_{kl} \varphi_0^k |\phi_2|^{2l}\ .
\ea
The terms of the Lagrangian~(\ref{Lnlo}) describe the rotations, and
the coupling between rotations and vibrations, while anharmonic
corrections to the vibrations are encoded in the
remainder~(\ref{Rnlo}). It is assumed that the ``wave function
renormalizations'' $Z_{00}, Z_{20}, Z_{02}, Z_{22}$ are of order
$\xi$. Under this assumption, the remainder $R_{\rm NLO}$ merely adds
perturbative corrections to the vibrations.  As we are mainly
interested in the lowest energetic vibrational states (the band heads)
and the modifications of the rotational bands due to the vibrational
coupling, we do not consider these anharmonicities.  A Legendre transformation
yields the Hamiltonian
\ba
H_{\rm NLO} &=& H_{\rm LO} 
+ {1\over 2 v^2}\left(p_\beta^2  +{1\over \sin^2\beta}
\left(p_\alpha -2 l_2 \cos\beta\right)^2\right) \ .
\ea
The comparison with the angular momentum~(\ref{q2}) shows that
\be
\label{HofQ}
H_{\rm NLO} = H_{\rm LO} + {1\over 2v^2}\left(Q^2 - 4l_2^2\right) \ .
\ee  
Note that the identification of the squared vacuum expectation value
$v^2$ with the moment of inertia is arbitrary but convenient. We could
also have employed the low-energy constant $C_0$ and avoided any reference
to $v$. The vacuum expectation value $v$ is not an observable, and the
most general Lagrangian can be expressed in terms of the variable $\varphi_0$
instead of $\phi_0=v+\varphi_0$. 

Let us turn to the quantization of the Hamiltonian at next-to-leading
order.  From the simple form of the classical Hamiltonian~(\ref{HofQ})
we expect a rotational band upon each vibrational state, with a spin
of the band-head equal to $2l_2$.  It is instructive to derive this
anticipated result.  For the quantization of the Nambu-Goldstone modes we use
Eq.~(\ref{momenta}). This yields the Hamiltonian operator
\ba
\label{hamvibrot}
\hat{H}_{\rm NLO} &=& \hat{H}_{\rm LO} 
- {1\over 2v^2}
{1\over \sin\beta}\partial_\beta\left(\sin\beta \partial_\beta\right) 
+{1\over 2v^2}{1\over\sin^2\beta}
\left(-i\partial_\alpha - 2\hat{l}_2\cos\beta\right)^2 \ .
\ea 
The eigenfunctions of $\hat{H}_{\rm NLO}-H_{\rm LO}$ are again given
in terms of the Wigner $d$ function as $e^{-im\alpha}d_{m
  l_2}^l(\beta)$.  The energies at order $\xi$ thus are
\ba
\label{Evibrot}
E(n_0, n_2, l_2, l)&=& \omega_0\left(n_0+{1\over 2}\right) 
+{\omega_2\over 2}(2n_2 + |l_2| +1) 
\nonumber\\
&+& {1\over 2v^2} \left(l(l+1)-(2l_2)^2\right)  \ .
\ea
Here, the angular momentum $l$ is an integer with $l\ge 2|l_2|$ and
depends on the $l_2$ value of the vibrational state with quantum
numbers $(n_0,n,l_2)$ under consideration. The
spectrum~(\ref{Evibrot}) is the well known rotational-vibrational
spectrum: upon each vibrational state, there is a rotational band.
Note that the moment of inertia is, at this order of the effective
theory, equal for each vibrational state.

The low-energy constants of the effective theory
are the vacuum expectation value $v$ (which fixes the moment of
inertia), and the frequencies $\omega_0$ and $\omega_2$ (which
determine the spacing of the ground state to so-called $\beta$ band
with quantum numbers $(n_0=1, n_2=0, l_2=0)$ and the so-called
$\gamma$ band with quantum numbers $(n_0=0, n_2=0,l_2=\pm 1)$,
respectively. If we are only interested in these three bands (as is
often the case), there is no need to determine the low-energy
constants of the anharmonic corrections~(\ref{Rnlo}).

\subsubsection{Next-to-next-to-leading order}

At next-to-next-to-leading order (N$^2$LO) (${\cal O}(\xi^2/\Omega)$), new terms
enter. There will be (relatively uninteresting) terms $R_{\rm N^2LO}$
that only depend on the vibrational degrees of freedom. More
interesting are the additional terms that couple rotations and
vibrations. These are
\ba
\Delta L_{\rm N^2LO}&=& 4|\phi_2|^2 E_z^2 \nonumber\\
&+&D_0\left(E_x^2+E_y^2\right)\varphi_0^2  
+F_0\left(E_x^2+E_y^2\right)\dot{\varphi}_0^2 \nonumber\\
&+& D_1 \varphi_0 \left(\phi_2 E_{-1}^2 +\phi_{-2} E_{+1}^2\right) 
+F_1\dot{\varphi}_0 \left(E_{+1}^2 D_t{\phi}_{-2} + E_{-1}^2 D_t{\phi}_2\right) \nonumber\\
&+& D_2\left(E_x^2+E_y^2\right)|\phi_2|^2
+F_2\left(E_x^2+E_y^2\right)|D_t{\phi}_2|^2 \ .
\ea
Here, $D_0$, $D_1$, and $D_2$ are dimensionless low-energy constants
of order one, while $F_1$, $F_2$, and $F_3$ are expected to be of
order $\Omega^{-2}$. Many of these terms are not considered in the 
generalizations~\cite{Gneuss,Hess1980,Greiner} of the Bohr Hamiltonian.

Let us briefly discuss the quantization procedure. As is well known,
the quantization of a classical Hamiltonian is only without ambiguity
in flat space, i.e.  when the metric exhibits no curvature. For a
metric with nonzero curvature, one can think of the system as a
constrained system, i.e. the motion is constrained to a (curved)
hypersurface in a higher-dimensional space. Kaplan, Maitra, and
Heller~\cite{Kaplan} showed that the quantization of such a
constrained system depends on the exact nature of the constraints,
i.e. on details regarding the implementation of the constraining
forces which -- in the limit of infinite forces -- confine the
motion to the hypersurface. These authors point out that the {\it
  physical} situation (i.e. the implementation of the constraints in a
limiting process) resolves the ambiguity. Our quantization follows
this rule. The leading order vibrational Hamiltonian is quantized
without ambiguity. When the rotations are coupled to the vibrations,
we know that -- in the limit of infinite frequencies $\omega_0$ and
$\omega_2$ -- the physics of Nambu-Goldstone bosons as presented in
Sect.~\ref{sec:NG} must result. Thus, we employ the previously derived
quantization rules. When considering terms beyond next-to-leading
order, the quantization rules remain unchanged.

\section{Coupling of fermions to Nambu-Goldstone modes}
\label{sec:fermions}

For deformed odd-mass nuclei, it is not sufficient to simply include
terms that are first order in the time derivative (such as Wess-Zumino
terms) into the Lagrangian. Such nuclei oft exhibit
several low-lying band heads with rotational bands upon them. In this
case, fermionic degrees of freedom must be included in the
description, and we need to coupled nucleon fields $N$ to
Nambu-Goldstone modes. As discussed in Section~\ref{sec:coset}, we
write
\be
N=g\chi \ ,
\ee
where $g\in {\cal H}=SO(2)$ as given in Eq.~(\ref{gEuler}). The most
general rotationally invariant Lagrangian consists of combination of
$\chi$, $D_t\chi$, and the functions $E_x$ and $E_y$ that is formally
invariant under rotations around the $z$-axis. In lowest order, we have
the Lagrangian~\cite{Wiese}
\ba
\label{fermionLag}
L &=& {C_0\over 2}\left(E_x^2 + E_y^2\right) + \chi^\dagger \hat{T}\chi 
+ \chi^\dagger \left(i\partial_t +E_z\hat{J}_z\right) \chi 
+C_1 \chi^\dagger \left( E_x\hat{J}_x+E_y\hat{J}_y\right) \chi \ .
\ea
Here, $\hat{T}$ is the operator of the kinetic energy (or the
single-particle energy) for the fermion.  The
Lagrangian~(\ref{fermionLag}) reminds us of the
particle-rotor model~\cite{BM,Greiner}. The choice of the spin $q$ of
the fermion field is determined by the nucleus under consideration. On
could, of course, identify $\chi$ with a two-spinor $\chi(\vec{r})$
(Then, we would also need to integrate over space in the Lagrangian.)
However, for heavy nuclei, it might be more adequate to identify
$\chi$ with the shell-model orbital $(n, l, j, \tau_z)$ (denoting
radial quantum number, orbital angular momentum, total spin, and
isospin projection, respectively) that is most relevant for the
open-shell nucleus under consideration.

It is insightful to consider the limit of a spin-$q$ fermion with
components $\chi=(\chi_q,\vec{0})^T$. Let us assume that $\chi_q$ is
the ground state of the operator $\hat{T}$, with $\hat{T}\sim \Lambda
\gg \xi$.  In this case, the low-energy Lagrangian~(\ref{fermionLag})
for the Nambu-Goldstone modes becomes (we use $|\chi_q|^2=1$)
\be
L = {C_0\over 2} \left(E_x^2 + E_y^2\right) + q E_z  
+ \langle\chi|\hat{T}|\chi\rangle\ .
\ee
Apart from the irrelevant constant $\langle\chi|\hat{T}|\chi\rangle$,
this Lagrangian has the same form as the Lagrangian~(\ref{nlag_univ})
with the Wess-Zumino term. This example shows how the Wess-Zumino term
arises from high-energy fermionic degrees of freedom.  The detailed
discussion of fermions coupled to Nambu-Goldstone modes is beyond the
scope of this paper. However, the techniques we developed in this
paper seem to be applicable to this case, too. For a nucleus with a
finite ground-state spin, one would need to couple the Nambu-Goldstone
modes to quadrupole bosons and single-particle degrees of freedom.

\section{Summary} 
\label{sec:summary}
This paper developed an effective theory for deformed nuclei. The
approach exploits the spontaneous breaking of the rotational symmetry
and is based on the separation of scale between low-energetic
Nambu-Goldstone rotational modes, higher energetic quadrupole modes,
and single-particle degrees of freedom at much higher energies. The
nonlinear realization of the rotation group is key to the effective
theory, and a power counting is established. We derived the Lagrangian
(and Hamiltonian) for the Nambu-Goldstone modes and the for the 
quadrupole bosons coupled to Nambu-Goldstone modes in next-to-leading
order. At this order in the power counting, well known results from
phenomenological models were rederived in a model-independent way.
More interesting phenomena are expected at next-to-next-to-leading
order, as the effective theory predicts the appearance of terms that
are not employed in the phenomenological models.  The effective theory
treats nuclei with finite spins in their ground states (odd-odd and
odd-mass nuclei) on equal footing to nuclei with zero ground-states
spins (even-even nuclei).

\section*{Acknowledgments}
  The author gratefully acknowledges discussions with R. J. Furnstahl,
  W. Nazarewicz, A.  Schwenk, and H. A. Weidenm{\"u}ller. He thanks
  the Institut f{\"u}r Kernphysik, Technische Universit{\"a}t
  Darmstadt, and the GSI Helmholtzzentrum f{\"u}r Schwerionenforschung
  for their hospitality.  This work was supported by the U.S.
  Department of Energy under Grants Nos. DE-FG02-07ER41529 and
  DE-FG02-96ER40963, and by the Alexander von Humboldt-Stiftung.

\appendix
\label{appendix}

\section{Legendre Transformation}
\label{app:legendre}
Let the Lagrangian depend on velocities $\partial_t\vec{x}$ ($\vec{x}=(x_1,\ldots,x_N)$) as follows
\be
L={1\over 2} \left(\partial_t\vec{x}\right)^T \hat{G}\partial_t\vec{x} +\vec{k}^T\partial_t\vec{x} \ .
\ee
Here $\hat{G}$ denotes the symmetric mass matrix, $^T$ denotes the
transpose, and $\vec{k}$ is a constant vector.  Then, the Legendre
transformation
\be
p_j={\partial L\over\partial \dot{x}_j} 
\ee
yields the Hamiltonian function
\be
H={1\over 2}\left(\vec{p}-\vec{k}\right)^T \hat{G}^{-1} \left(\vec{p}-\vec{k}\right) \ .
\ee 

For the quantization of the Nambu-Goldstone modes $(\alpha,\beta)$, 
we follow~\cite{Podolsky} and write the
Lagrangian  as a quadratic form
\ba 
L={1\over 2} (\dot\beta,\dot\alpha) \, \hat{G} \, 
\left(\begin{array}{c}\dot\beta \\ \dot\alpha\end{array}\right) \ . 
\ea 
This defines the ($2\times 2$) matrix of the metric $\hat{G}$. The Hamiltonian
becomes (let us use units where $\hbar=1$)
\ba 
\hat{H}&=& {-1\over 2\sqrt{{\rm det}\hat{G}}}
(\partial_\beta,\partial_\alpha) \hat{G}^{-1} \sqrt{{\rm det}
  \hat{G}} \left(\begin{array}{c}\partial_\beta\\
    \partial_\alpha\end{array}\right) \ .  
\ea 
For an in-depth discussion of the quantization of constrained systems,
the reader is refered to Ref.~\cite{Kaplan}.

\section{Nambu-Goldstone modes revisited}
\label{app:ng}
The derivations presented in Sect.~\ref{sec:coset} are more formal
than necessary if one is only interested in the physics of
Nambu-Goldstone modes. Here, we follow Leutwyler~\cite{Leut93} for a
quicker derivation. The dynamics of the Nambu-Goldstone modes is
determined by the coset $SO(3)/SO(2)$ which is isomorphic to the
two-sphere $S^2$. Thus, the Nambu-Goldstone modes parameterize the
two-sphere, and we can choose the parameterization
\be 
\label{nv}
\vec{n}(\alpha,\beta) =\left(\begin{array}{c}
\cos{\alpha}\sin{\beta}\\\sin{\alpha}\sin{\beta}\\\cos{\beta}\end{array}\right) \ .
\ee 
Here, $\alpha$ and $\beta$ are time-dependent variables. The simplest
low-energy Lagrangian that is invariant under time reversal is
\ba
\label{Ln}
L&=&{C_0\over 2} \left(\partial_t\vec{n}\right)\cdot
\left(\partial_t\vec{n}\right) 
= {C_0\over 2} \left(\dot{\beta}^2 + \dot{\alpha}^2\sin^2\beta\right) \ .
\ea
This is Eq.~(\ref{nlag}). One can pursue this direction further and
also construct the Wess-Zumino term. Details are given in
Ref.~\cite{Leut93}.

\section{Transformation properties under rotations}
\label{app:trafo}
Let
\be
\label{gapp}
g(\alpha,\beta) = e^{-i\alpha \hat{J}_z}e^{-i\beta \hat{J}_y}
\ee
denote an element of the coset $SO(3)/SO(2)$ and  
\be
\label{rapp}
r(\alpha,\beta,\gamma) = e^{-i\alpha \hat{J}_z}e^{-i\beta \hat{J}_y}e^{-i\gamma \hat{J}_z} 
\ee
be a general rotation parameterized in terms of the three Euler angles. Let
\be
\label{happ}
h(\gamma)=e^{-i\gamma \hat{J}_z}
\ee
be a rotation of the subgroup ${\cal H}=SO(2)$. 

The product $rg$ is again a rotation, and we have
\be
r(\alpha_2,\beta_2,\gamma_2)g(\alpha,\beta) =
r(\tilde{\alpha},\tilde{\beta},\tilde{\gamma})\ .
\ee
Expressions for $(\tilde{\alpha},\tilde{\beta},\tilde{\gamma})$ in terms of 
the other angles are well known~\cite{bible}
\ba
\label{trafo_angles}
\cot(\tilde{\alpha}-\alpha_2)&=& \cos\beta_2 \cot(\alpha+\gamma_2)
+\cot\beta {\sin\beta_2\over\sin(\alpha+\gamma_2)} \ , \nonumber\\
\cos\tilde{\beta} &=& \cos\beta\cos\beta_2 
-\sin\beta\sin\beta_2 \cos(\alpha+\gamma_2) \ , \nonumber\\
\cot\tilde{\gamma}&=& \cos\beta \cot(\alpha+\gamma_2)
+\cot\beta_2 {\sin\beta\over\sin(\alpha+\gamma_2)} \ . 
\ea
Due to the definitions~(\ref{gapp})-(\ref{happ}), we also have
\be
r(\tilde{\alpha},\tilde{\beta},\tilde{\gamma}) = g(\tilde{\alpha},\tilde{\beta})h(\tilde{\gamma}) \ . 
\ee
Under a rotation $r(\alpha_2,\beta_2,\gamma_2)$ the element $g$ of the coset
transforms as
\ba
g=g(\alpha,\beta)\to \tilde{g}\equiv g(\tilde{\alpha},\tilde{\beta}) \ , 
\ea
and $\tilde{g}=\tilde{g}(g,r)$ depends on the Nambu-Goldstone modes
$(\alpha,\beta)$ that parameterize $g$ and the rotation angles
$(\alpha_2,\beta_2,\gamma_2)$ of $r$. The rotation thus maps the
Nambu-Goldstone modes $(\alpha,\beta)$ into
$(\tilde{\alpha},\tilde{\beta})$, and one needs to employ the
transformations~(\ref{trafo_angles}) to obtain explicit results. These
transformation laws are complicated, but we do not need them for the
construction of Lagrangians that are invariant under rotations (as
shown in Subsection~\ref{sec:coset}). They do, however, enter the
derivation of the conserved quantities, i.e. the components of the
angular momentum. For this purpose, we need to know the transformation laws
for rotations by infinitesimal angles $\delta\omega_k$ around the  $k=x,y,z$ axes. 
The corresponding rotations are 
$r(-\pi/2,\delta\omega_x,\pi/2)$, $r(0,\delta\omega_y,0)$, and
$r(\delta\omega_z,0,0)$ respectively. Employing the transformation
laws~(\ref{trafo_angles}) we find
\ba
\left(\begin{array}{c}
\delta\alpha\\
\delta\beta\end{array}\right)
=\hat{M}
\left(\begin{array}{c}
\delta\omega_x\\
\delta\omega_y\\
\delta\omega_z\end{array}\right) \ ,
\ea
with 
\ba
\hat{M}=
\left(\begin{array}{ccc}
-\cot\beta \cos\alpha & -\cot\beta \sin\alpha & 1 \\
-\sin\alpha & \cos\alpha & 0
\end{array}\right) \ .
\ea
This is Eq.~(\ref{trafo_euler}). 

We can repeat these considerations for the quadrupole phonons.  
As shown in Sect.~\ref{sec:coset},
under rotations $r(\alpha_2,\beta_2,\gamma_2)$, the quadrupole modes
$\phi$ transform as
\be
\label{phitrans}
\phi\to h(\tilde{\gamma}) \phi = e^{-i\tilde{\gamma}\hat{J}_z}\phi \ .
\ee 
According to the transformation laws~(\ref{trafo_angles}), under 
infinitesimal rotations by $\delta \omega$ around the $x$, $y$,
and $z$ axis, the angle $\tilde{\gamma}$ becomes $\tilde{\gamma}=
\delta\omega {\cos\alpha\over\sin\beta}$, $\tilde{\gamma}=\delta\omega
{\sin\alpha\over\sin\beta}$, and $\tilde{\gamma}=0$ respectively.
The application of Eq.~(\ref{phitrans}) shows that the 
quadrupole fields (with $\phi_2=\phi_{2r}+i\phi_{2i}$) transform as
\ba
\left(\begin{array}{c}
\delta\phi_{2r}\\
\delta\phi_{2i}\\
\delta\varphi_0\end{array}\right) 
=\hat{N}
\left(\begin{array}{c}
\delta\omega_x\\
\delta\omega_y\\
\delta\omega_z\end{array}\right) \ , 
\ea
where
\ba
\hat{N} \equiv 
\left(\begin{array}{ccc}
+2{\cos\alpha\over\sin\beta}\phi_{2i} &+2{\sin\alpha\over\sin\beta}\phi_{2i} & 0\\
-2{\cos\alpha\over\sin\beta}\phi_{2r} &-2{\sin\alpha\over\sin\beta}\phi_{2r} & 0\\
0 & 0 & 0
\end{array}\right) \ .
\ea  
As expected for the nonlinear realization, $\varphi_0$ is invariant under rotations. 
We can use these transformation laws to apply Noether's theorem.

\section{Application of Noether's theorem}
\label{app:noether}
Let us recall Noether's theorem for our purposes. Consider a
Lagrangian $L$ of coordinates $q_\nu$ and velocities $\dot{q}_\nu$,
$\nu=1,\ldots,N$.  Let us also consider a coordinate transformation
that depends on parameters $\omega_k$, $k=1, \ldots, K$.  
For infinitesimal small arguments $\delta \omega_k$ the coordinates change by
\be
\delta q_\nu = \sum_{k=1}^K \hat{M}_{\nu k} \delta \omega_k  \ .
\ee  
The coordinate transformation changes the Lagrangian by the amount 
\be
\delta L = \sum_{\nu=1}^N \left(
{\partial L\over\partial \dot{q}_\nu}\delta\dot{q}_\nu +{\partial L\over\partial q_\nu}\delta q_\nu \right) \ . 
\ee
We employ the equations of motions
\be
{\partial L\over\partial q_\nu}=\partial_t{\partial L\over\partial \dot{q}_\nu} \ , \quad\mbox{$\nu=1,\ldots,N$} \ , 
\ee
and find
\ba
\delta L = \partial_t\sum_{\nu=1}^N {\partial L\over \partial \dot{q}_\nu}\delta q_\nu \ .
\ea
Let us consider the case that only one parameter $\omega_k$ is varied. 
Thus, the Lagrangian changes by  
\be
\label{deltaL}
\delta L = \delta \omega_k  
\partial_t\sum_{\nu=1}^N {\partial L\over \partial \dot{q}_\nu}\hat{M}_{\nu k} \ .
\ee
In our case, the transformations are rotations. In the case that the
ground state is invariant under time reversal, we have $\delta L=0$
for arbitrary $\delta\omega_k$. Thus,
\ba
\label{conservedNG}
Q_k&\equiv& \sum_{\nu=1}^N {\partial L\over \partial \dot{q}_\nu}\hat{M}_{\nu k}
=\sum_{\nu=1}^N p_\nu\hat{M}_{\nu k}
\ea 
is a conserved quantity (i.e. $\partial_t Q_k=0$).  Here, $p_\nu$ is
the momentum conjugate to $q_\nu$.  We insert the matrix $\hat{M}$ 
into Eq.~(\ref{conservedNG}) and obtain the three components of the
angular momentum~(\ref{conserved_NG}).

We can repeat these considerations when quadrupole phonons are coupled
to the Nambu-Goldstone modes. The application of Noether's theorem  yields
\ba
\left(\begin{array}{c}
Q_x\\
Q_y\\
Q_z\end{array}\right) 
=
\hat{M}^T
\left(\begin{array}{c}
p_\alpha \\
p_\beta\end{array}\right)
+\hat{N}^T 
\left(\begin{array}{c}
p_{2r}\\
p_{2i}\\
p_0\end{array}\right) \ .
\ea 
Thus ($l_2\equiv \phi_{2r}p_{2i}-\phi_{2i}p_{2r}$),  
\ba
Q_x &=& -2{\cos\alpha\over\sin\beta}l_2
-p_\beta \sin\alpha -p_\alpha \cot\beta \cos\alpha \ , \nonumber\\
Q_y &=& -2{\sin\alpha\over\sin\beta}l_2
+p_\beta \cos\alpha -p_\alpha \cot\beta \sin\alpha \ , \nonumber\\
Q_z &=& p_\alpha \ .
\ea  
This is Eq.~(\ref{conserved_phi}).

In the case that the ground state breaks time reversal symmetry, 
the Lagrangian is not invariant under
infinitesimal rotations but changes by the amount~(\ref{delta_LWZ}) due to the Wess-Zumino term. 
Thus, 
\ba
\label{deltaLWZ}
\delta L &=&  \delta L_{\rm WZ}
= q \left( \delta \omega_x \partial_t\left({\cos\alpha\over\sin\beta}\right) 
+ \delta \omega_y \partial_t\left({\sin\alpha\over\sin\beta}\right) \right) \ .
\ea
This change clearly is a total time derivative, and equating the
expressions~(\ref{deltaL}) and (\ref{deltaLWZ}) for $k=x, y, z$ yields the conserved
quantities~(\ref{conserved_WZ}).

\section{Solution of differential equation}
\label{app:dgl}
We discuss the diagonalization of the Hamiltonians~(\ref{ham_LO}) and (\ref{hamvibrot}). The Nambu-Goldstone modes are essentially governed by the Hamiltonian
\ba
\label{diffeq}
\hat{H}&=&-{1\over \sin\beta}\partial_\beta\sin\beta \partial_\beta 
+{1\over \sin^2\beta}\left(-i\partial_\alpha+q\cos{\beta}\right)^2 \ .
\ea
For the eigenfunction $f_{mq}^l(\alpha,\beta)$, we make the ansatz 
\be
f_{mq}^l(\alpha,\beta)=g_{mq}^l(\beta) e^{-im\alpha} \ . 
\ee
Here $l$ is a quantum number that needs to be determined.
This yields the eigenvalue equation
\ba
\left\{-{1\over \sin\beta}\partial_\beta\sin\beta \partial_\beta 
+{1\over \sin^2\beta}\left(m-q\cos{\beta}\right)^2 \right\} g_{mq}(\beta) 
= E(l,m,q) g_{mq}(\beta) \ .
\ea
We expand the square, rewrite $\cos^2\beta=1-\sin^2\beta$ and find
\ba
\left\{-{1\over \sin\beta}\partial_\beta\sin\beta \partial_\beta 
+{1\over \sin^2\beta}\left(m^2-2mq\cos{\beta}+q^2\right) \right\} g_{mq}^l(\beta) 
\nonumber\\
= \left(E(l,m,q)+q^2\right) g_{mq}^l(\beta) \ .
\ea
For the eigenfunctions of this differential operator, we recall that the
Wigner $D$ functions 
\be
\label{WignerD}
D_{m q}^l(\alpha,\beta,\gamma) \equiv e^{-im\alpha} d_{m q}^l(\beta) e^{-iq\gamma}
\ee
(and the functions $d_{mq}^l(\beta)$ themselves) 
solve the differential equation~\cite{bible}
\ba
\label{WignerDdiff}
\left\{-{1\over\sin\beta}\partial_\beta\left(\sin\beta \partial_\beta\right) 
+ {1\over \sin^2\beta}\left(m^2-2mq\cos\beta +q^2\right)\right\} D_{mq}^l(\alpha,\beta,\gamma)
\nonumber\\
= l(l+1)D_{mq}^l(\alpha,\beta,\gamma) \ . 
\ea
Thus, we have to identify the eigenfunctions as $g_{mq}^l(\beta) =
d_{mq}^l(\beta)$, and the eigenvalue is $E(l,m,q)=l(l+1)-q^2$.  It is
well known~\cite{bible} that for $q=0$, the Wigner function
$D_{m0}^l(\alpha,\beta,\gamma)$ is proportional to the spherical
harmonics $Y_{l-m}(\beta,\alpha)$.

\end{document}